\documentclass[journal]{IEEEtran}
\usepackage{amsmath,amsfonts}
\usepackage{algorithmic}
\usepackage{algorithm}
\usepackage{array}
\usepackage[caption=false,font=normalsize,labelfont=sf,textfont=sf]{subfig}
\usepackage{textcomp}
\usepackage{stfloats}
\usepackage{url}
\usepackage{verbatim}
\usepackage{graphicx}
\usepackage[version=4]{mhchem}
\usepackage{hyperref}
\usepackage{authblk}
\usepackage{cite}
\usepackage{xcolor}
\usepackage{multirow}
\def\BibTeX{{\rm B\kern-.05em{\sc i\kern-.025em b}\kern-.08em
    T\kern-.1667em\lower.7ex\hbox{E}\kern-.125emX}}
\usepackage{balance}
\begin{document}
\title{Nanosecond-scale discrete wavelength switching in feedback-controlled single-gain-section multi-wavelength lasers}
\author{Mathieu Ladouce, Pablo Marin-Palomo, and Martin Virte
%\author[1]{Pablo, Marin-Palomo}
%\author[1]{Martin Virte}% <-this % stops a space
%\thanks{Affiliation and acknowledgments}% <-this % stops a space
\thanks{This work was supported by the European Research Council through Starting Grant COLOR'UP No. 948129 (M.V.), the Research Foundation Flanders through Grant No. G0G0319N (M.V.) and postdoctoral fellowship Grant No. 1275924N (P.M.-P.), and the METHUSALEM program of the Flemish Government.}
\thanks{Mathieu Ladouce, Pablo Marin-Palomo, and Martin Virte are with Brussels Photonics Team (B-PHOT), Vrije Universiteit Brussel (VUB), 1050 Brussels, Belgium (e-mail: Mathieu.Ladouce@vub.be; Pablo.Marin-Palomo@vub.be; Martin.Virte@vub.be).}
\thanks{The data that supports the findings of this article are openly available \cite{ladouce_2026_18317452}.}
%\thanks{Manuscript received April 19, 2021; revised August 16, 2021.}
}


\maketitle

\begin{abstract}
We investigate discrete wavelength switching in single-gain-section multi-wavelength lasers monolithically integrated on InP with phase-controlled optical-feedback.
By modulating the feedback phase, nanosecond-scale wavelength switching is experimentally demonstrated with transition times below 2.5 ns. Measurements consistently show that the switching time decreases with stronger optical feedback and larger phase-modulation amplitudes.
Transitions from lower to higher modal gain are faster. 
We support the experimental observations with a multi-mode extension of the Lang–Kobayashi rate-equation model.
We analyze the influence of laser, feedback-cavity, and modulation parameters on the switching dynamics, and highlight the role of mode coupling.
These results highlight the potential of integrated multi-wavelength lasers for compact and high-speed all-optical networking systems.
\end{abstract}


\section{Introduction}
\IEEEPARstart{O}{n-chip} lasers with agile wavelength control are of strong interest for applications including differential absorption LiDAR \cite{Iwai2021}, data center interconnects \cite{Zhou2023}, optical switching networks \cite{Marin-Palomo2025}, micro- and mm-wave photonics \cite{Akin2025}, and interferometry \cite{Houairi2009}. Depending on the use case, this agility is pursued either through continuous sweep or switch among discrete wavelengths, with switching times of a few nanoseconds or below considered highly desirable \cite{Ueda2020, Coldren2022}.
\\
A wide variety of integrated approaches has therefore been explored.
Hybrid platforms that exploit the Pockels effect in lithium niobate (LiNbO3), have demonstrated exceptionally high tuning speeds at petahertz-per-second  rate \cite{Li2022, Snigirev2023, Siddharth2025}, though with limited wavelength tuning range.
On the other side, several devices have reported sub-nanosecond discrete switching \cite{Ueda2020,Wang2025A, Wang2025B}, but they rely on multiple active sections and effectively operate as independent coupled lasers.
Fully monolithic solutions based on phase tuning in sampled-grating distributed Bragg reflectors (SG-DBR) \cite{Yu2002, Simsarian2006}, or Vernier selection using cascaded ring resonators \cite{Segawa2007, Watanabe2015,Verolet2018}, provide nanosecond-scale discrete tuning, with spans ranging from tens of picometers to several nanometers.
However, these architectures are typically optimized for single-wavelength operation and do not readily support simultaneous multi-wavelength emission, which is attractive for functions such as photonic heterodyning for THz generation \cite{Akin2025} or wavelength routing \cite{Marin-Palomo2025}.
\par
Multi-wavelength lasers (MWLs) are integrated laser diodes capable of emitting at several longitudinal modes from a single active medium, either simultaneously or individually, in a controllable manner.
In contrast with other tunable laser structures, MWLs use a fundamentally different wavelength-selection mechanism: all longitudinal modes are coupled since they share the same gain medium, and the lasing wavelengths are determined by controlled interference within a feedback path.
Their ability to provide multiple, coupled, and THz-spaced optical carriers within a compact monolithic device makes them attractive for applications such as dynamic wavelength channel allocations in future optical networks, wavelength conversion, or spectroscopy \cite{Pawlus2019, Pawlus2023, Virte2023, Abdollahi2024}.
Understanding how rapidly they can reconfigure their emission becomes, therefore, essential.
\\
In the MWL devices used in this work \cite{Pawlus2019, Pawlus2023, Virte2023, Abdollahi2024}, tuning relies exclusively on the electro-optic control of the round-trip phase in a monolithically integrated feedback cavity.
Adjusting this single parameter modifies the interference conditions that govern mode competition inside the main cavity, enabling deterministic selection of one or several lasing modes.
% We investigate the temporal dynamics of such wavelength switching in detail.
We experimentally report nanosecond-scale switching between lasing wavelengths, with measured transition times below 2.5 ns.
We introduce a measurement procedure for evaluating wavelength-transition times and employ both experimental characterization and a multi-mode extension of the Lang–Kobayashi rate-equation model to investigate the impact of device parameters and operating conditions on the switching speed.
Our experimental results highlight differences between rising and falling edges of the transition, the robustness of the laser response to imperfect phase-modulation waveforms, the diminution of switching time with increased phase-modulation amplitudes, and the limited sensitivity to modulation ramp time and modal gain differences.
Numerical exploration of the parameter space confirms the key roles of modulation amplitude and feedback strength, and clarifies the impact of the modal gain difference.
Overall, we demonstrate that optical-feedback-based wavelength switching can reach nanosecond timescales without requiring high-speed electronic devices, e.g., with bandwidths $\geq 1$ GHz, offering a promising route toward compact and energy-efficient all-optical wavelength routing.

\section{MWL and experimental setup\label{sec:2-ExperimentalSetup}}
\begin{figure*}[ht]
    \centering
    \includegraphics[width=1\linewidth]{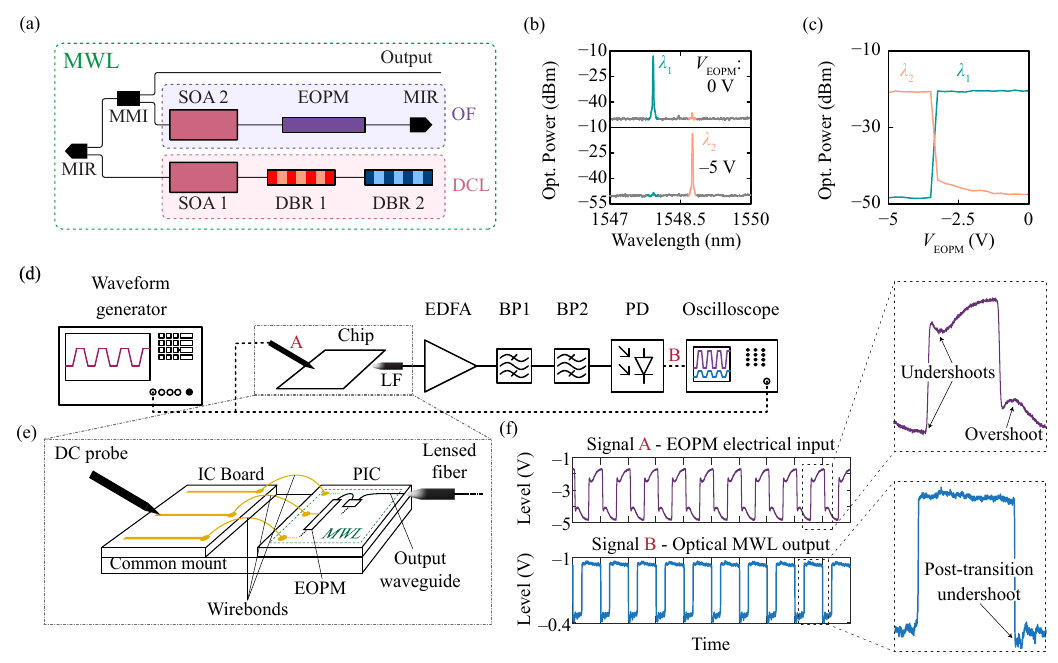}
    \caption{
    Wavelength switching in feedback-controlled on-chip multi-wavelength lasers.
    (a) Schematic of the MWL integrated circuit: the dual-cavity laser (DCL) is delimited by a multimode interference reflector (MIR) on one side and by two distributed Bragg reflectors (DBRs) on the other side and includes a semiconductor optical amplifier (SOA 1) constituting the single gain medium.
    The optical feedback cavity, connected with the DCL via a 1 $\times$ 2 multimode interference coupler (MMI), contains SOA 2 and an electro-optic phase modulator (EOPM) for feedback strength and round-trip phase control, respectively.
    (b) By applying $V_\text{EOPM}=0$ V or $V_\text{EOPM}=-5$ V, we observe the single emission at $\lambda_1=1547.9$ nm or $\lambda_2=1548.7$ nm, respectively.
    (c) Output power of the MWL at $\lambda_1$ and $\lambda_2$; we observe a sharp switch between the two wavelengths at $V_\text{EOPM} \approx -3.4 V$.
    (d) Experimental setup for switching time measurements.
    Two signals are recorded via the scope: the driving modulation at the DC probe level (signal A) and the photocurrent generated by the output light from the MWL at $\lambda_1$ impinging into the photodiode (PD) (signal B).
    (e) Schematic of the electrical connections at the chip-level for modulation of the EOPM. (f) Representative temporal responses of the phase modulation at the probe level (signal A) and the laser response (signal B) with zoomed views on the overshoot and undershoot features around wavelength transitions.
    }
    \label{fig:1-setup}
\end{figure*}

The multi-wavelength laser is integrated on an InP generic platform from SmartPhotonics, see the schematic of the MWL in Fig. \ref{fig:1-setup}(a).
Each active component on the photonic integrated circuit (PIC) is connected to an interconnect (IC) board via wirebonds.
The chip is mounted so that all active components share a common ground. The MWL consists of a dual-cavity laser formed with two distributed Bragg reflectors (DBR) – DBR 1 and DBR 2 with Bragg wavelengths at 1548 nm and 1537 nm, respectively – and a semiconductor optical amplifier (SOA 1) acting as the single gain medium.
The dual-cavity laser is monolithically integrated with a feedback cavity in which the strength can be adjusted with another amplifier (SOA 2) and the feedback round-trip phase is controlled by applying a voltage to an electro-optic phase modulator (EOPM) \cite{Pawlus2019, Pawlus2023}.
The light emitted by the MWL is collected through a lensed fiber (LF).
The measured optical spectrum contains two families of modes associated with the reflectivity bandwidths of each DBR.
Those bandwidths can be shifted towards lower wavelengths by injecting current into DBRs.
A controlled change of the voltage $V_\text{EOPM}$ applied to the EOPM in the feedback cavity can trigger a switch of the emitted wavelength \cite{Pawlus2023}, as pictured in Fig. \ref{fig:1-setup}(b-c).
The experimental setup described in Fig. \ref{fig:1-setup}(d) is used to repeatedly trigger wavelength switching and record the response of the MWL.
In order to apply a periodic signal to the EOPM, we use a DC probe to contact the metal pad in the IC board that corresponds to the EOPM, see Fig. \ref{fig:1-setup}(e).
The electrical input signal is generated either by a function generator (Agilent 33500B, 30 MHz, 250 MSa/s) or an arbitrary waveform generator (Keysight M8194A, 45 GHz, 120 GSa/s), depending on the desired ramp time $t_\text{ramp}$ between the lower and upper voltage levels.
Additional practical details on the feedback phase modulation are provided in Appendix \ref{sec:AppendixA-ModulationSignal}.
High frequency RF interconnects were not intended in the design process of the MWL, which is equipped with DC metal tracks.
Therefore, impedance mismatching between the waveform generator, the DC probe and the metal tracks is expected.
We observe the presence of overshoots and undershoots in the modulated signal measured before the probe, see Fig. \ref{fig:1-setup}(f).
The signal from the laser is amplified with an erbium-doped fiber amplifier (EDFA), filtered by two optical bandpass filters to isolate a given wavelength for analysis, and then measured with a photodiode (New Focus 1554-B, 12 GHz) connected to an oscilloscope (Keysight MDO4104, 1 GHz, 5GSa/s).
Despite the arguable quality of the modulation, the response of the laser is close to a square waveform. 

The wavelength switching time is defined as the duration required for the laser to switch between two wavelengths in response to a controlled change of the feedback phase in the cavity.
In both experiment and simulation, wavelength switching is triggered by applying a periodic trapezoidal modulation to the feedback phase, producing successive transitions between two stable lasing states.
Experimentally, due to the small difference between the two wavelengths, only 0.8 nm, and limited rejection at the bandpass filter edges, we estimate the side-mode suppression ratio to be between 6 and 10 dB.
As a result, there is a clear residual contribution of the other mode, which may distort the lower-state level and introduce fluctuations in the measured waveform around transitions.

\section{Nanosecond wavelength switching performance}
\begin{figure}
    \centering
    \includegraphics[width=1\linewidth]{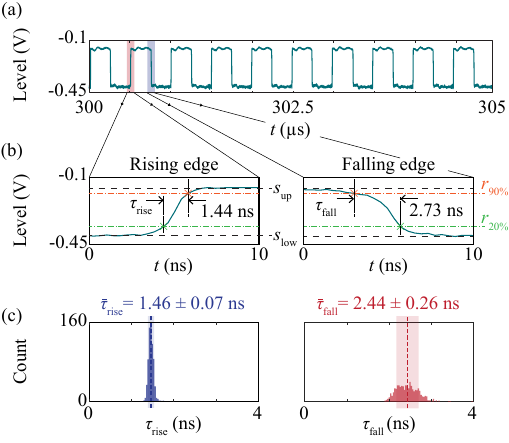}
    \caption{
    (a) Representative temporal response of the laser under trapezoidal modulation of the feedback phase at a frequency of 2 MHz.
    (b) Detailed view of the wavelength switching transitions.
    The MWL optical output is oscillating between two steady-state levels: the upper state level $s_\text{up}$, corresponding to the emission at $\lambda_1 = 1547.9$, and the lower state level $s_\text{low}$, corresponding to the suppression of $\lambda_1$ being replaced by the emission at $\lambda_2 = 1548.7$ nm.
    For each transition edge, the switching time is measured as the elapsed time between the two reference levels $r_{20\%}$ and $r_{90\%}$.
    (c) Corresponding distribution of the measured rise (blue) and fall (red) switching times extracted from the complete waveform presented in panel (a).
    }
    \label{fig:2-SwitchingTimeExample}
\end{figure}

Fig. \ref{fig:2-SwitchingTimeExample}(a) shows a representative temporal waveform of the detected optical power under trapezoidal phase modulation of the EOPM.
The laser exhibits clean transitions between its two emission states, with well-defined upper and lower levels.
To accurately quantify switching times, the waveform is segmented into sets of 20 transitions [see the zoom on the rising and falling edges Fig. \ref{fig:2-SwitchingTimeExample}(a)], for which the upper and lower state levels are determined using a histogram-based method \cite{IEEE181-2011} [see Fig. \ref{fig:2-SwitchingTimeExample}(b)].
Because undershoots artificially depress the apparent lower-state level, the standard 10–90 \% transition-time definition would overestimate the true switching duration.
To avoid this bias, we define fixed reference levels at 20 \% and 90 \% of the amplitude difference between the steady-state levels.
The rise and fall switching times, denoted $\tau_\text{rise}$ and $\tau_\text{fall}$, are then computed as the time intervals between these reference levels and averaged over all transitions.
It is important to note that the measured switching times don't include the switching reaction time of the laser, i.e., the time delay between the instant the variation in $V_\text{EOPM}$ is applied to the EOPM and the lasers emission exhibits a change in wavelength.
The distribution of rise and fall times extracted from repeated switching events is shown in Fig. \ref{fig:2-SwitchingTimeExample}(c).
The rising and falling transitions exhibit different average durations, respectively $1.46\pm0.07$ and $2.44\pm0.26$ ns, indicating an intrinsic asymmetry between the two switching directions.
Assuming a Gaussian-equivalent bandwidth-limited response, the intrinsic rise-time limit of a 1 GHz oscilloscope is approximately 0.34 ns \cite{Brown1992}, which is well below the switching times measured here.
This estimate, however, only accounts for the oscilloscope front-end response; additional bandwidth limitations from the photodiode, amplifier, and interconnects may further increase the minimum resolvable switching time.

\begin{figure}
    \centering
    \includegraphics[width=1\linewidth]{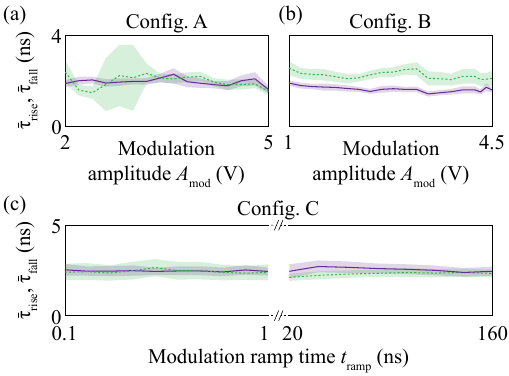}
    \caption{
    Rise (purple) and fall (green, dashed line) switching time averages as a function of the modulation amplitudes (a,b) and ramp times (c).
    The effect of $t_\text{ramp}$ was evaluated over two ranges with different orders of magnitude.
    Colored areas correspond to the standard deviation of the switching times.
    Details on the operational parameters of configuration A, B and C are provided in Table \ref{tab:1-Configurations}.
    }
    \label{fig:3-ExpModulationCondInfluence}
\end{figure}

To evaluate how the modulation waveform affects the wavelength switching speed, we experimentally vary two key parameters of the EOPM drive signal: the modulation amplitude $A_\text{mod}$ and the linear ramp time $t_\text{ramp}$ between the lower and upper levers of the trapezoidal waveform.
The corresponding switching times are summarized in Fig. \ref{fig:3-ExpModulationCondInfluence}.
Panels (a) and (b) show the dependence of the mean switching rise and fall times on the modulation amplitude for two distinct configurations of laser and feedback parameters, respectively labeled Config. A and Config. B.
Each investigated configuration of driving parameters is described in Table \ref{tab:1-Configurations}.
\begin{table*}[]
\caption{
List of investigated configurations with their corresponding operational parameters and the obtained extreme switching rise and fall times average values.
\label{tab:1-Configurations}
}
\begin{center}
\begin{tabular}{|c|l|l|l|l|l|l|}
\hline
\multirow{8}{*}{\textit{\textbf{Parameters}}} &
  \textbf{Configuration} &
  \multicolumn{1}{c|}{\textbf{A}} &
  \multicolumn{1}{c|}{\textbf{B}} &
  \multicolumn{1}{c|}{\textbf{C}} &
  \multicolumn{1}{c|}{\textbf{D}} &
  \multicolumn{1}{c|}{\textbf{E}} \\ \cline{2-7} 
 & Temperature                       & 16 °C             & 16 °C               & 16 °C                & 16 °C                   & 16 °C                     \\ \cline{2-7} 
 & SOA 1 current                     & 75 mA             & 75 mA               & 70 mA                & 70 mA                   & 70 mA                     \\ \cline{2-7} 
 & SOA 2 current                     & 23 mA             & 33 mA               & 22 mA                & 26 mA                   & \textbf{28.5 to 33.75 mA} \\ \cline{2-7} 
 & DBR 1 current                     & 9.5 mA            & 9.5 mA              & 9.3 mA               & \textbf{9.5 to 11.6 mA} & 9.0 mA                    \\ \cline{2-7} 
 & DBR 2 current                     & 11 mA             & 11 mA               & 5.0 mA               & 11.7 mA                 & 9.0 mA                    \\ \cline{2-7} 
 & $A_\text{mod}$                    & \textbf{2 to 5 V} & \textbf{1 to 4.5 V} & 2.1 V                & 1.8 V                   & 2.5 V                     \\ \cline{2-7} 
 & $t_\text{ramp}$                   & 8.4 ns            & 8.4 ns              & \textbf{0.1 to 160 ns} & 8.4 ns                  & 8.4 ns                    \\ \hline\hline
\multirow{4}{*}{\textit{\textbf{Results}}} &
  Smallest $\bar{\tau}_\text{rise}$ &
  1.60 ns &
  1.41 ns &
  2.38 ns &
  2.04 ns &
  \textbf{1.39 ns} \\ \cline{2-7} 
 & Largest $\bar{\tau}_\text{rise}$  & 2.28 ns           & 1.81 ns             & 2.72 ns              & \textbf{3.49 ns}        & 2.10 ns                   \\ \cline{2-7} 
 & Smallest $\bar{\tau}_\text{fall}$ & \textbf{1.47 ns}  & 2.04 ns             & 2.12 ns              & 2.27 ns                 & 2.34 ns                   \\ \cline{2-7} 
 & Largest $\bar{\tau}_\text{fall}$  & \textbf{2.37 ns}  & 2.55 ns             & 2.65 ns              & 3.35 ns                 & 3.03 ns                   \\ \hline
\end{tabular}
\end{center}
\end{table*}
Panel (c) presents the influence of the modulation ramp time over two different ranges.
In the first configuration A, the modulation amplitude does not seem to influence the switching time.
However, it appears that, in the second configuration B, larger modulation amplitudes yield slightly shorter switching times, indicating that stronger phase excursions accelerate the mode transition.
The ramp time of the modulation signal has little to no influence on the measured switching time, even when the electrical edge is significantly slower than the intrinsic optical transition.
This confirms that high-speed electronic drivers, e.g., with bandwidths $\geq 1$ GHz, are not required to achieve nanosecond switching in this system.
We note that the examinable range of modulation amplitudes $A_\text{mod}$ is limited firstly by the maximum save voltage of the EOPM, and secondly by the recurrence of switching when shifting continuously the round-trip feedback phase; a too large modulation amplitude would trigger unwanted wavelength switching.

\begin{figure}
    \centering
    \includegraphics[width=1\linewidth]{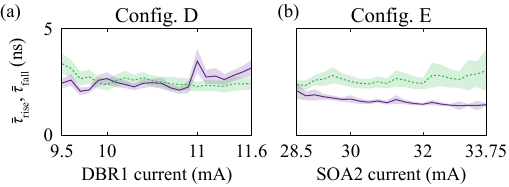}
    \caption{
    Rise (purple) and fall (green, dashed line) switching time averages as a function of the DBR 1 (a) and SOA 2 (b).
    Colored areas correspond to the associated standard deviation of the switching times.
    Details on the operational parameters of configuration D and E are provided in Table \ref{tab:1-Configurations}.
    }
    \label{fig:4-ExpParameterInfluence}
\end{figure}

To explore how intrinsic and operational parameters affect the switching dynamics, we adopt a two-step approach: first, we vary the laser parameters while keeping the optical-feedback conditions fixed; second, we vary the OF parameters while maintaining a constant laser configuration.
The presented configurations D and E (see details in Table\ref{tab:1-Configurations}) are selected arbitrarily to provide a relatively large range of variation of the parameter under test in which wavelength switching is possible; outside the investigated variation ranges, switching was not observed, and for other configurations, the variation range was eventually shorter.
We first investigate the impact of modal gain asymmetry on the switching time by adjusting the current injected into DBR 1.
When injecting current into a DBR, its Bragg wavelength, and thus its reflectivity bandwidth shifts towards smaller wavelengths.
Therefore the gain experienced by the modes increases or decreases depending on their wavelength.
Fig. \ref{fig:4-ExpParameterInfluence}(a) presents the experimental results of switching times as a function of DBR 1 current.
Across the tested range, no clear dependency is observed, suggesting that the measured switching times are largely insensitive to small variations in modal gain.

Next, we assess the role of the feedback strength by varying the current injected into SOA 2 in a fixed laser configuration.
The resulting switching times are shown in Fig. \ref{fig:4-ExpParameterInfluence}(b).
While the influence is relatively small, a consistent trend emerges: stronger feedback slightly reduces the wavelength switching time.
This observation confirms that the feedback cavity contributes to the dynamics, and optimizing its strength can be a viable strategy to enhance switching speed.

\section{Numerical analysis}
We have experimentally demonstrated that the ramp times during wavelength switching events remain relatively robust to operational conditions.
However, to examine whether the observed trends are inherent to the dynamics of the laser-feedback system or dependent on parameter configurations, we rely on numerical analysis to explore regimes beyond the experimental reach.

To perform our numerical analysis, we use the multi-mode extension of the Lang-Kobayashi rate equations, which is the “Model B” introduced in \cite{Koryukin2004} and already presented in \cite{Ladouce2026}.
This model, considered for two lasing modes --- $\lambda_1$ and $\lambda_2$ --- only, describes the time evolution of the field amplitudes and carrier densities.
A cross-saturation parameter $\beta$ is introduced in the carrier equations as a phenomenological coupling between the modes, ranging from $\beta=0$ for independent modes to $\beta=1$ for fully coupled modes.
Modal gain terms $g_1$ and $g_2$ capture the wavelength-dependency of the gain and will typically be expressed in terms of modal gain difference $\Delta g = g_1 - g_2$.
Regarding feedback parameters, the strength and delay are denoted as $\kappa$ and $\tau_\text{OF}$, respectively.
In this study, similarly to the experimental conditions, the modal feedback phases $\Psi_1$ and $\Psi_2$ are modulated in time with a periodic trapezoidal function that is characterized by a feedback phase amplitude $A_\text{mod}$, a feedback phase offset $\phi_\text{mod}$ that typically corresponds to the switching point, a frequency $f$ and a linear ramp time between lower and upper levels $t_\text{ramp}$.
The feedback phases of each mode are constantly separated by $\pi$: $\Psi_2(t) = \Psi_1(t) + \pi$.

The remaining fixed parameters of the model are indicated in Table \ref{tab:2-FixedParameters}.
Rate equations are normalized with respect to the photon lifetime but, to provide results in direct comparison with the experiment, the switching time values are given in unnormalized units, assuming a photon lifetime $\tau_\text{p}=3$ ps.
Details on the normalization are explicitly stated in \cite{Koryukin2004}.
Note that the purpose of using this model is not providing quantitative predictions but trends regarding the influence of operational parameters on the switching times.

\begin{table}
    \begin{center}
    \caption{List of simulation parameters described in \cite{Ladouce2026} with their respective fixed values.}
    \label{tab:2-FixedParameters}
    \begin{tabular}{|c|c|}
        \hline
        Linewidth enhancement factor $\alpha$ & 3\\
        \hline
        Spontaneous emission noise factor $\beta_\text{sp}$ & $10^{-12}$\\
        \hline
        Carrier to photon lifetime ratio $T$ & 1000\\
        \hline
        Normalized injection parameter $P$ & 0.5\\
        \hline
        Self-saturation parameter $\beta_{mm}$ & 1\\
        \hline
    \end{tabular}
    \end{center}
\end{table}

\begin{figure}
    \centering
    \includegraphics[width=1\linewidth]{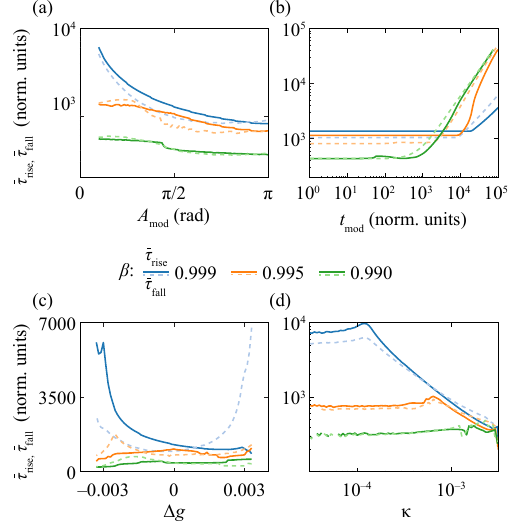}
    \caption{
    Rise (solid lines) and fall (dashed lines) switching time averages derived numerically from the time traces of the first mode for three values of cross-saturation parameter $\beta$, as a function of the: (a) phase modulation amplitude $A_\text{mod}$, (b) phase modulation ramp time $t_\text{ramp}$, (c) modal gain difference $\Delta g$ and (d) feedback strength $\kappa$.
    Unless modified explicitly, simulation parameters are: $\Delta g = 0$, $\kappa=0.001$, $\tau_\text{OF}=200$, $A_\text{mod}=\pi$, $\phi_\text{mod}=\pi/2$, $t_\text{ramp}=10$.
    }
    \label{fig:5-SimCurves}
\end{figure}

Figure \ref{fig:5-SimCurves} presents the numerically computed switching times as a function of phase-modulation amplitude (panel a), modulation ramp time (panel b), modal gain difference $\Delta g$ (panel c), and feedback strength $\kappa$ (panel d).
For each panel, results are shown for three values of the cross-saturation parameter $\beta$ to enable direct comparison with the experimental trends.
\\
The simulations reproduce all qualitative tendencies observed experimentally.
Increasing the modulation amplitude yields faster transitions.
This effect is more visible with larger values of $\beta$.
The dependence on the modulation ramp time is also consistent with the measurements: the switching time remains nearly constant for fast to moderately slow modulation ramp times.
Until the modulation becomes slow enough that the laser dynamics begin to follow the imposed waveform.
In this regime, the switching time grows approximately linearly with the modulation ramp time.
The $\beta$ value determines the edge-time threshold at which this transition occurs, with smaller $\beta$ enabling faster switching.

\begin{figure}
    \centering
    \includegraphics[width=1\linewidth]{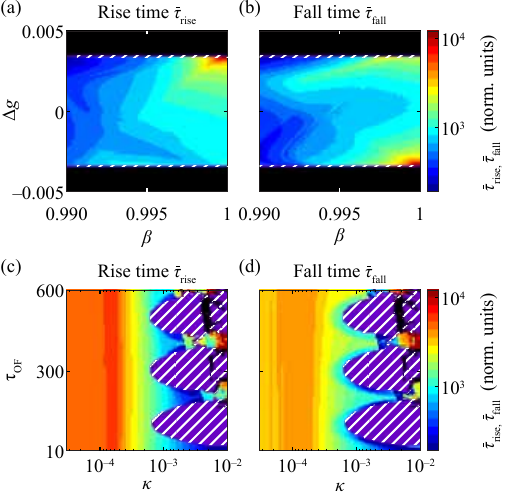}
    \caption{
    Maps showing the switching rise (left) and fall (right) times, derived numerically from the time trace of the first mode, as a function of: (a,b) cross-saturation parameter $\beta$ and modal gain difference $\Delta g$, and (c,d) feedback strength $\kappa$ and delay $\tau_\text{OF}$.
    Non-switching configurations are displayed in black while regions where the switching time could not be evaluate due to the presence of dynamics are represented as purple with white dashed areas.
   Unless modified explicitly, simulation parameters are: $\Delta g = 0$, $\kappa=0.001$, $\tau_\text{OF}=200$, $A_\text{mod}=\pi$, $\phi_\text{mod}=\pi/2$, $t_\text{ramp}=10$.
    }
    \label{fig:6-SimMaps}
\end{figure}

The influence of modal gain asymmetry follows the expected physical intuition: switching toward the mode with the larger gain is faster than switching toward the gain-penalized mode.
However, this effect becomes clearly visible only for $\beta$ values close to the unity such as $\beta=0.999$ in the given example.
For smaller $\beta$ values, the switching times decreases but the curves become nearly constant, mirroring the weak gain-asymmetry dependence observed experimentally.
Finally, the feedback-strength dependence reveals that the switching time stays essentially constant over a broad range of $\kappa$ before decreasing once the feedback becomes sufficiently strong.
This threshold shifts toward higher $\kappa$ when $\beta$ is small, which indicates once more that reduced cross-saturation leads to intrinsically faster transitions.
Overall, the simulations suggest that the relative robustness observed experimentally—across modulation amplitudes, ramp times, modal gains, and feedback strengths—can be explained by an experimental value of the cross-saturation $\beta$ that is relatively small.
\\
To further clarify how the laser–feedback system behaves beyond experimentally accessible regimes, we perform a parameter-space exploration.
Figure \ref{fig:6-SimMaps}(a,b) shows the combined influence of modal gain difference $\Delta g$ and cross-saturation $\beta$, while Fig. \ref{fig:6-SimMaps}(c,d) explores the space defined by feedback strength $\kappa$ and delay $\tau_\text{OF}$.

The $\Delta g$--$\beta$ map confirms the intrinsic asymmetry between rise and fall transitions: a non-zero gain difference speeds up switching in one direction while slowing it down in the other.
Larger $\beta$ amplifies this asymmetry, whereas smaller $\beta$ produces faster switching overall and reduces the sensitivity to $\Delta g$.
These results also illustrate a trade-off between achieving a large switching-suppression ratio—favored by higher $\beta$ and smaller $\Delta g$, as demonstrated in \cite{Ladouce2026}—--and maintaining short switching times.
\\
The $\kappa$--$\tau_\text{OF}$ map confirms that stronger feedback produces faster wavelength transitions while indicating they are largely insensitive to the feedback delay.
The regions where a reduction of switching time appears are not primarily linked to the value of $\tau$ itself but instead coincide with the onset of dynamical regimes introduced by the feedback.
This observation suggests that, within the stable operational regime used experimentally, the delay plays only a minor role in determining the switching speed.

\section{Conclusion}
We have demonstrated that wavelength switching driven by phase-tuned optical feedback in a multi-wavelength semiconductor laser enables nanosecond-scale transitions, with switching times as low as 2.5 ns.
The measurements reveal a remarkable robustness to phase modulation conditions: the square response of the laser remains relatively consistent against imperfect driving waveform, and the switching time is largely insensitive to variations in modulation ramp time.
At the same time, the switching speed improves slightly with increased modulation amplitude.
As a result, high-speed electronic drivers, e.g., with bandwidths $\geq 1$ GHz, are not required to reach the nanosecond regime.

The switching time is reduced when transitioning toward the gain-favored mode, though this effect is subtle in the experiment.
Numerical analysis suggests that this limited sensitivity originates from a relatively small effective cross-saturation parameter $\beta$.
Reducing $\beta$ systematically shortens the switching time but also weakens the side-mode suppression ratio, highlighting a fundamental trade-off between speed and spectral selectivity.
\\
Simulations further confirm that feedback strength has only a moderate influence on the switching time, while the feedback delay plays a negligible role except in regimes where it induces dynamical behavior.
Still, we believe that further studies may investigate devices with shorter feedback cavities and increased feedback strengths.
More broadly, the laser and feedback parameters primarily determine whether the device can sustain wavelength switching, but within the operational range, they have a limited impact on the actual switching time.	
\\
Overall, the combined experimental and numerical results establish phase-controlled feedback as a simple and robust strategy for nanosecond wavelength switching in integrated multi-wavelength lasers, well-suited for applications requiring fast, reconfigurable optical carriers.

\appendix
\label{sec:AppendixA-ModulationSignal}
To trigger switching between two wavelengths of the MWL repeatedly, we modulate the feedback phase with a periodic trapezoidal signal.
%, as represented in Fig.\ref{fig:7-TrapSignal}.
%\begin{figure}
%    \centering
%    \includegraphics[width=1\linewidth]{Figures/Figure07.pdf}
%    \caption{
%    Graphical representation of the parameters of the trapezoidal modulation of the feedback round-trip phase: the modulation frequency $f$ , the modulation amplitude $A_\text{mod}$, the modulation offset $\phi_\text{mod}$ and the linear transition time between the lower and upper levels $t_\text{ramp}$.
%    }
%    \label{fig:7-TrapSignal}
%\end{figure}
The upper and lower levels of the modulation signal should correspond to different single-wavelength emission states of the MWL, e.g., the upper state corresponds to the single emission at $\lambda_1$ and suppressed emission at $\lambda_2$.
\\
Experimentally, we vary the feedback phase by applying a voltage to the EOPM in the feedback cavity.
Therefore, the level, the offset $\phi_\text{mod}$, and the amplitude $A_\text{mod}$ of the feedback phase modulation signal are expressed in volts, while they are directly expressed in radians in the context of numerical investigations.
In practice, the offset $\phi_\text{mod}$ is determined in advance by measuring at which tension $V_\text{EOPM}$ the switch between the two wavelengths occurs.
For example, for the configuration presented in Fig.\ref{fig:1-setup}, the switch happens around $V_\text{EOPM}\approx-3.4\text{ V}$ [see panel (a)] so that we set the feedback phase modulation offset at the same level: $\phi_\text{mod}=-3.4\text{ V}$ [see panel (f)].
The modulation amplitude $A_\text{mod}$ can take any value provided that the modulation signal level remains negative to prevent forward bias of the EOPM.
Two different electrical signal generators were employed depending on the experimental requirements.
In most measurements, we used a function generator (Agilent 33500B, 30 MHz, 250 MSa/s) that allows direct control of both the modulation amplitude $A_\text{mod}$ and the offset $\phi_\text{mod}$ over a wide voltage range ($-5$ to $+5$ V).
However, its limited analog bandwidth results in a relatively large minimum modulation ramp time $t_\text{ramp}$ of 8.4 ns.
For experiments specifically investigating the influence of $t_\text{ramp}$ on the switching time, i.e. results presented in Fig.\ref{fig:3-ExpModulationCondInfluence}(c,d), we instead used an arbitrary waveform generator (Keysight M8194A, 45 GHz, 120 GSa/s), which provides $t_\text{ramp}$ as low as 0.1 ns.
This comes at the cost of reduced output voltage range and the absence of a configurable DC offset.
Consequently, an external bias tee (Mini-Circuits ZFBT-6GW) and RF amplifier (iXblue DR-AN-40-MO, 40 GHz) were inserted in the circuitry to achieve the desired modulation levels.

%\section{Experimental results compilation}
%All experimental results of the switching rise and fall times measurements presented in this work are compiled into a single histogram shown in Fig.\ref{fig:8-CompilationHistogram}.

%\begin{figure}
%    \centering
%    \includegraphics[width=1\linewidth]{Figures/Figure08.pdf}
%    \caption{
%    Normalized distribution of the measured switching rise (left) and fall (right) times accross all experiments presented in this work.
%    The switching time occurrences were normalized by the total number of sample in each investigated configuration.
%    These numbers are provided in Table \ref{tab:1-Configurations}.
%    }
%    \label{fig:8-CompilationHistogram}
%\end{figure}

\bibliographystyle{IEEEtran}
\bibliography{references}

\end{document}